\documentclass[aps,prb,preprint,groupedaddress]{revtex4}

\usepackage{epsfig}
\begin{document}
\title{

Predicting interface structures: From SrTiO$_3$ to graphene

}

 \author{Georg Schusteritsch}
 \email{g.schusteritsch@ucl.ac.uk}
 \affiliation{
Department of Physics and Astronomy, University College London, Gower Street, London WC1E 6BT, United Kingdom \\
}
 \author{Chris J. Pickard }
 
  \email{c.pickard@ucl.ac.uk}
\affiliation{
Department of Physics and Astronomy, University College London, Gower Street, London WC1E 6BT, United Kingdom \\
}
\affiliation{
London Institute for Mathematical Sciences, 35a South Street, Mayfair, London W1K 2XF, United Kingdom\\
}

\date{\today}

\begin{abstract}

We present here a fully first-principles method for predicting the atomic structure of interfaces. Our method is based on the {\it ab initio} random structure searching (AIRSS) approach, applied here to treat two dimensional defects. The method relies on repeatedly generating random structures in the vicinity of the interface and relaxing them within the framework of density functional theory (DFT). The method is simple, requiring only a small set of parameters that can be easily connected to the chemistry of the system of interest, and efficient, ideally adapted to high-throughput first-principles calculations on modern parallel architectures. Being first-principles, our method is transferable, an important requirement for a generic computational method for the determination of the structure of interfaces. Results for two structurally and chemically very different interfaces are presented here, grain boundaries in graphene and grain boundaries in strontium titanate (SrTiO$_3$). We successfully find a previously unknown low energy grain boundary structure for the graphene system, as well as recover the previously known higher energy structures. For the SrTiO$_3$ system we study both stoichiometric and non-stoichiometric compositions near the grain boundary and find previously unknown low energy structures for all stoichiometries. We predict that these low energy structures have long-range distortions to the ground state crystal structure emanating into the bulk from the interface.

 \end{abstract}

\maketitle

\section{Introduction}

Determining the atomic structure of interfaces is a problem of great importance in many areas of physics and materials science. Interfaces strongly influence the mechanical and electronic properties of most polycrystalline materials and play a crucial role for heterostructures. Our understanding of the structure of interfaces at an atomic level and how it relates to the physical properties of the bulk material is however still very limited and an area of active research. Though significant improvements in experimental imaging and image analysis methods have been made, it is still exceedingly difficult to uniquely determine the atomic structure of interfaces experimentally and theoretical methods are often necessary to supplement experimental results.~\cite{Finnis:Sutton:STO:review:2010, STO:GB:Ikuhara:PRB:2011} On the other hand, significant improvements in theoretical first-principles methods for the prediction of the ground state crystal structures of particularly bulk materials have been made.~\cite{Ma:Perspective:2014, AIRSS:Review:Pickard:Needs:2011} By applying such first-principles methods to interfaces, a reversed approach, where one first predicts the structure of interfaces theoretically and determines their properties from theory, without the need of prior experimental results, is within the realm of possibility. Once the atomic structure is known theoretically it would then in principle also be possible to connect it to experimental results by generating simulated HRTEM images~\cite{SimulatedMicroscopy} and EELS spectra~\cite{Pickard:EELS:2009, Pickard:EELS:1997}. The ability to altogether independently predict the atomic structure of interfaces using theoretical methods would enable us to better understand the relation of interfaces to the physical properties of materials, which in turn paves the way to develop materials with unique interfaces that give them particular properties.

The structure prediction of interfaces is a great challenge; any theoretical method to tackle this problem has to be able to reliably and accurately describe the atomic structure and be highly transferable so that it can be applied to a wide variety of materials systems. At the same time the method should be efficient enough to be able to predict the crystal structure for a sufficiently large region surrounding the interface. A limited number of computational approaches to predict the ground state of interfaces based on evolutionary algorithms and basin-hopping have been proposed.~\cite{GA:STO:GB:Finnis:Sutton:2010, DE:Graphene:GB:Gong, Basin:Hopping:Wei:2009} These were however either based on searching with classical interatomic potentials,~\cite{GA:STO:GB:Finnis:Sutton:2010, DE:Graphene:GB:Gong} thereby lacking the transferability and accuracy of first-principles approaches, or in the case of Ref.~\onlinecite{Basin:Hopping:Wei:2009}, which used a basin-hopping approach in combination with density functional theory (DFT), only a single atomic layer at the interface was addressed without considering non-stoichiometric conditions. A method that allows for efficient treatment of variable stoichiometries is however of crucial importance especially for many of the technologically important complex oxides.~\cite{GA:STO:GB:Finnis:Sutton:2010, Benedek:STO:GB:2012}

We address this challenging theoretical problem using a fully first-principles structure prediction method that is sufficiently efficient to consider a large region surrounding the interface for both stoichiometric and non-stoichiometric conditions. To our knowledge a fully first-principles structure prediction study of interfaces or grain boundaries with variable stoichiometry has not been attempted previously. The approach we take is that of {\it ab initio} random structure searching (AIRSS).~\cite{AIRSS:PRL:Pickard:Needs:2006} This has previously been successfully used for the prediction of bulk crystal structures~\cite{AIRSS:Aluminium:CJP:2010, AIRSS:Ammonia:CJP:2008, AIRSS:Hydrogen:CJP:2007, AIRSS:Oxygen:CJP:2012} and point defect structures.~\cite{AIRSS:Zintl:defect:CJP:2013, AIRSS:HNO:Si:CJP:2009, AIRSS:H:Si:CJP:2008} We apply this method here to treat grain boundaries, an important subset of interfaces. To illustrate the broad applicability and transferability of our approach, we present a study of grain boundaries in two structurally and chemically very different materials: graphene and strontium titanate (SrTiO$_3$).

The effect of grain boundaries on the electronic and mechanical properties of graphene has been studied extensively using theoretical methods.~\cite{DE:Graphene:GB:Gong, Graphene:Louie:2010, Graphene:Zhao:2012, Graphene:Yakobson:2010, Graphene:Pantelides:2011, Graphene:Luo:2011, Graphene:Lu:2012} Much of this work has been based on structures from molecular dynamics (MD) or were created by inspection using intuition. Previous work on the prediction of bulk materials however suggests that this approach rarely leads to low energy structures.~\cite{AIRSS:PRL:Pickard:Needs:2006}  We choose here to study a grain boundary with a tilt angle of $\theta=30^{\circ}$ between the two grains, equivalent to an interface between an armchair- and zigzag-terminated grain. This type of grain boundary can be found experimentally.~\cite{Graphene:Exp:GB:Ruoff} Its physical properties have been the subject of several recent theoretical studies~\cite{DE:Graphene:GB:Gong, Graphene:Louie:2010, Graphene:Zhao:2012, Graphene:Yakobson:2010, Graphene:Pantelides:2011, Graphene:Luo:2011, Graphene:Lu:2012} and it has also been studied using a differential evolution algorithm in combination with interatomic potentials.~\cite{DE:Graphene:GB:Gong} Using our approach we find a previously unknown low energy structure in addition to the already known structures studied by Liu {\it et al.}~\cite{Graphene:Yakobson:2010} and Li {\it et al}.~\cite{DE:Graphene:GB:Gong}

For the SrTiO$_3$ system we consider a $\Sigma3\left(111\right)$ grain boundary. This and similar high-angle symmetric tilt grain boundaries have been studied experimentally~\cite{STO:GB:Ruehle:2003, STO:GB:Ikuhara:PRB:2011, STO:GB:Pennycook:2001} and theoretically.~\cite{STO:GB:Pennycook:2001, STO:GB:Cockayne:2010, STO:BTO:GB:Ikuhara:PRB:2008, STO:GB:Ikuhara:PRB:2011, Benedek:Chua:Finnis:2008} They were found to introduce unique electronic properties to the bulk~\cite{STO:GB:Pennycook:2001} and Uberuaga {\it et al.} showed that defect segregation can vary significantly depending on the atomic structure and stoichiometry of the grain boundary.~\cite{Benedek:STO:GB:2012} Chua {\it et al.} have studied two symmetric tilt grain boundaries using a genetic algorithm.~\cite{GA:STO:GB:Finnis:Sutton:2010}  This landmark work, addressing these very complex grain boundaries for variable stoichiometry, found several low energy equilibrium structures. However, their search algorithm relied on an interatomic potential, which, while computationally efficient, may result in inaccuracies especially for nonstoichiometric conditions or when resolving the rich set of low energy crystal phases of complex oxides is of importance. Our approach does not suffer from these restrictions in the same manner and we find several lower energy structures. Crucially, our searching method finds structures in lower energy phases near the grain boundary than in the work by Chua {\it et al}.~\cite{GA:STO:GB:Finnis:Sutton:2010} Our structures are found to include long-range phase distortions that emanate from the grain boundary, resulting from the specific geometry of the two grains and how their lattices are matched at the boundary. We  find that our approach is not sensitive to the initial crystal phases, another important aspect for reliably studying the structure of complex oxides. 

The remainder of the paper is organized as follows. In Sec.~\ref{Sec:Comp:Details} we summarize the random structure approach used to study the interfaces and the computational details of the DFT calculations for each system can be found. We discuss our results for the graphene and SrTiO$_3$ grain boundary systems in Secs.~\ref{Sec:Graphene} and \ref{Sec:STO}, respectively, and conclude in Sec.~\ref{Sec:Conclusion}.  

\section{Computational Details}\label{Sec:Comp:Details}

The {\it ab initio} random structure searching (AIRSS) method relies on placing atoms at random, yet physically sensible, positions followed by a rigorous structural optimization using DFT, in our case using CASTEP.~\cite{CASTEP} To treat grain boundaries we define a randomization region that separates two grains, illustrated for graphene and SrTiO$_3$ in Figs.~\ref{Fig:Graphene:GB:structures} and~\ref{Fig:STO:GB:structures}, respectively. The geometry of the two grains surrounding the randomization region determines the type of interface.  Although only grain boundaries are studied in this work, generalization to heterostructure interfaces and surfaces is in principle straightforward. Constraints are imposed to ensure that high energy and hence very unphysical structures are eliminated prior to geometry optimization. This primarily takes the form of imposing minimal interatomic distances for the initially random positions of the atoms in the randomization region. Several hundred random structures are generated for each stoichiometry and atomic density and, after geometry optimization, ranked according to their energy.

For both the SrTiO$_3$ and the graphene study we initially search with coarse parameters and soft pseudopotentials and then refine for our final calculations. Structural relaxations for both systems are performed using the Broyden-Fletcher-Goldfarb-Shanno (BFGS) method. We initially relax all forces to a magnitude of less than $0.05~\text{eV/\AA}$ for searching followed by a stricter tolerance of $0.01~\text{eV/\AA}$ for the final results. For the SrTiO$_3$ calculations the exact XC energy is approximated by the local density approximation (LDA).~\cite{Perdew:LDA:PRB:1981} Previous work suggests employing LDA for this specific system may be advantageous (see Ref.~\onlinecite{Benedek:Chua:Finnis:2008} and references therein) and has been used successfully by Chua {\it et al.} in their work on the prediction of the SrTiO$_3$ grain boundary using a genetic algorithm as the method to rank their final low energy interfaces.~\cite{GA:STO:GB:Finnis:Sutton:2010} The interactions between the valence electrons and the ionic cores are described using utrasoft pseudopotentials. For searching we use pseudopotentials that treat the valence electrons for the $3d^2$, $4s^2$ states for Ti, the $4s^2$, $4p^6$, $5s^2$ states for Sr and the $2s^2$ $2p^4$ states for O. Our final results are calculated using harder core-corrected on-the-fly-generated pseudopotentials that treat the $3s^2$ $3p^6$ $3d^2$ $4s^2$ states for Ti, the $4s^2$ $4p^6$ $5s^2$ states for Sr and the $2s^2$ $2p^4$ states for O. A plane wave cutoff energy of  $360~\text{eV}$ for searching and $520~\text{eV}$ for the final calculations was chosen to satisfy convergence. We use a Monkhorst-Pack mesh of $2\times2\times1$ for searching and $4\times4\times1$ for the final calculations. 
 We optimize the lattice vector perpendicular to the grain boundary for all final calculations but found it sufficient to search with fixed volumes for a given density in the randomization region. 

For graphene the exact XC energy is approximated by the PBE generalized gradient approximation.~\cite{Perdew:PRL77:1996} We have also performed separate searches using the LDA; this did not change the conclusions or order of the stability of the interfaces. We found it sufficient to use $\Gamma$-point calculations. All calculations were performed using ultrasoft pseudopotentials with the valence states 2s$^2$ 2p$^2$ for C, where a harder on-the-fly generated pseudopotential was used for the final results. The calculations for searching were performed using a plane wave energy cutoff of 280eV, whilst our final calculations used a cutoff of 500eV.

\section{Graphene Grain Boundary}\label{Sec:Graphene}

\begin{figure}
\center{
\includegraphics[width=8.6cm]{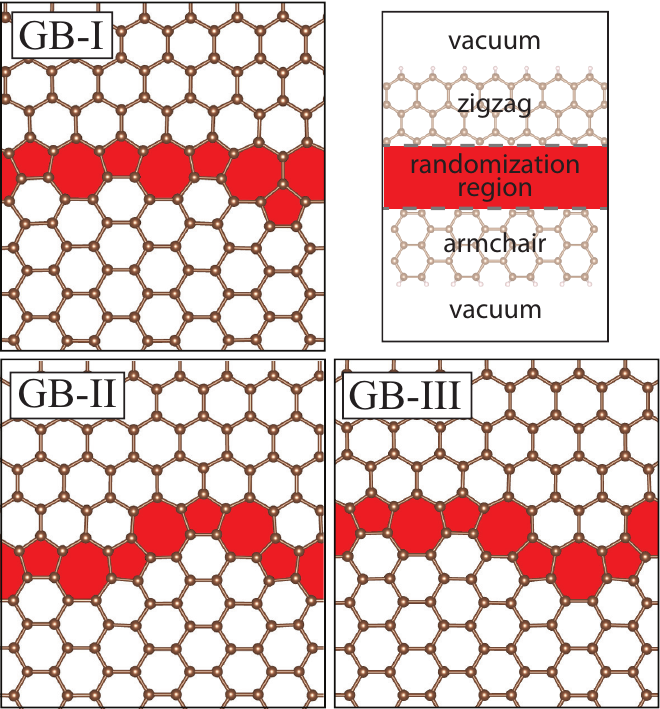}

\caption{
Graphene grain boundary structures between armchair and zigzag regions. The three lowest energy grain boundary structures are shown, with both GB-II and GB-III significantly lower in energy than GB-I. Also shown is the setup for searching for interface structures using AIRSS.
}
\label{Fig:Graphene:GB:structures}
}
\end{figure}

We first discuss our results for the graphene zigzag/armchair grain boundary. The setup for the calculations is shown in Fig.~\ref{Fig:Graphene:GB:structures}. The initial distance between the armchair and zigzag region is $3.0~\text{\AA}$. To minimize strain in the bulk away from the grain boundary a $\left(7,0\right)|\left(4,4\right)$ type geometry was chosen, resulting in a lattice mismatch of $1.0\%$. Previous studies often concentrated on a $\left(5,0\right)|\left(3,3\right)$ interface, with a significantly higher lattice mismatch of almost $4\%$. This strain is artificial, due to the periodic boundary conditions, and its effect should be minimized for searching. The length of the cell in the plane of the graphene sheet perpendicular to the grain boundary was chosen to be $20\text{~\AA}$ and a vacuum of $6.8\text{~\AA}$ perpendicular to the graphene sheet was found sufficient for searching; the latter was increased to $20\text{~\AA}$ for the final calculations of the interface energy. The length, L, parallel to the GB was chosen to be that of either the ideal armchair or zigzag region ($L_{ac}=17.07~\text{\AA}$ or $L_{zz}=17.25~\text{\AA}$). The size of the cell was held fixed during searching. The outer edges of the armchair and zigzag regions are terminated with hydrogen (H) and the only ionic constraint imposed during searching was that all H atoms at the armchair region were held fixed. This allowed for shear parallel to the grain boundary and also perpendicular displacements between the two grains. The interface energy, $\sigma$, for each grain boundary structure is determined using larger unit cells with 186 atoms and periodic boundary conditions perpendicular and parallel to the grain boundary. We define the interface energy, $\sigma$, for the graphene grain boundary in the usual manner:
 \begin{equation}\label{Eq:Graphene:interface}
\sigma=\frac{1}{2L}\left(G_{\text{tot}}-n_{\text{C}}\mu_{\text{C}}\right), 
\end{equation}
where G$_{\text{ tot}}$ is the Gibbs free energy of the cell containing the grain boundary, $n_{C}$ is the total number of carbon (C) atoms in the cell, $L$ is the length of the cell parallel to the grain boundary and $\mu_{C}$ is the chemical potential of C based on a calculation for ideal graphene.

Searching involved adding $N$ C atoms into the randomization region at random positions. We have studied the system for different numbers of C atoms and found our lowest energy structure for $N=7$ and $15$. This number of C atoms allows for formation of pentagons and heptagons across the grain boundary. We restrict ourselves to flat graphene sheets here and show the low energy structures in Fig.~\ref{Fig:Graphene:GB:structures}, with their respective interface energies summarized in Table~\ref{Table:graphene:GB:Ef}. The structure labeled GB-I had been assumed to be the lowest energy structure in most previous studies of the physical properties of the zigzag/armchair grain boundary. We find it to have an interface energy similar to previous work,~\cite{Graphene:Yakobson:2010, DE:Graphene:GB:Gong} but significantly higher in energy in comparison to structures GB-II and GB-III. Recent work by Li {\it et al.} also finds structure GB-II, with similar interface energy as in our work.~\cite{DE:Graphene:GB:Gong} We have found a new low energy structure, labeled in Fig.~\ref{Fig:Graphene:GB:structures} as GB-III.

The structure of GB-III is similar to GB-II, both consisting of alternating pentagons and heptagons as opposed to the ``fly-head" pattern of GB-I.~\cite{Graphene:Yakobson:2010} The periodicity of these heptagons and pentagons is however very different: GB-II consists of two heptagons on each side, whereas GB-III has three heptagons on one side with just one heptagon on the other side. This may hint at even lower energy structures for larger system sizes parallel to the grain boundary. The interface energy crucially depends on the cell length parallel to the grain boundary interface. This is not a well defined quantity for calculations using periodic boundary conditions parallel to the interface, since the simulated bulk above and below the grain boundary interface should have different lattice constants. There is therefore an inherent uncertainty in the interface energy given here and we include the interface energy for two cases where $L$ is set by either the armchair or zigzag region. We see that assuming $L$ matched for the armchair region results in the interface energy of GB-III to be lower than that of GB-II, while $L$ matched for the zigzag region reverses the order. To resolve this issue of the energetic order of the two types of grain boundary structures, one would need to increase the system size parallel to the grain boundary in order to appropriately reduce the artificial strain in the system. The $\left(7,0\right)|\left(4,4\right)$  type grain boundary used in the work here has a lattice mismatch of $1.0\%$. The next larger interface to lower this artificial strain build-up is a $\left(19,0\right)|\left(11,11\right)$ type grain boundary with a lattice mismatch of $0.3\%$. This structure has however a length, $L$, parallel to the grain boundary of more than $46$~\AA, making it unfeasibly large for conventional DFT calculations.

Our method allows us to quickly find the low energy structures for each $N$: All new and previously known structures shown in Fig.~\ref{Fig:Graphene:GB:structures} could be found multiple times with $N=15$ for less than 300 initial structures.

\begin{table}
 \begin{center}
  \begin{tabular*}{8.6cm}{@{\extracolsep{\fill}}l  c  c }
  \hline \hline
 		& $\sigma\left(L_{ac}\right)$ [eV/nm]		&$\sigma\left(L_{zz}\right)$ [eV/nm]		\\ \hline
 GB-I		& $4.28$								& $4.39$								\\
 GB-II	& $3.33$								& $3.18$								\\
 GB-III	& $3.29$								& $3.24$								\\
  \hline \hline
  \end{tabular*}
  \caption{
Grain boundary interface energies for the three lowest-energy grain boundary structures between zigzag and armchair graphene. Structure GB-III is found to be lower in energy than GB-II when the unit cell is constrained to the optimal length of an armchair cell ($L=L_{ac}$), whilst the order reverses when $L$ is constrained to be optimal for the zigzag region ($L=L_{zz}$).
   }
   \label{Table:graphene:GB:Ef}
 \end{center}
\end{table}

\section{Strontium Titanate Grain Boundary}\label{Sec:STO}

\begin{figure*}
\center{
\includegraphics[width=17.2cm, clip]{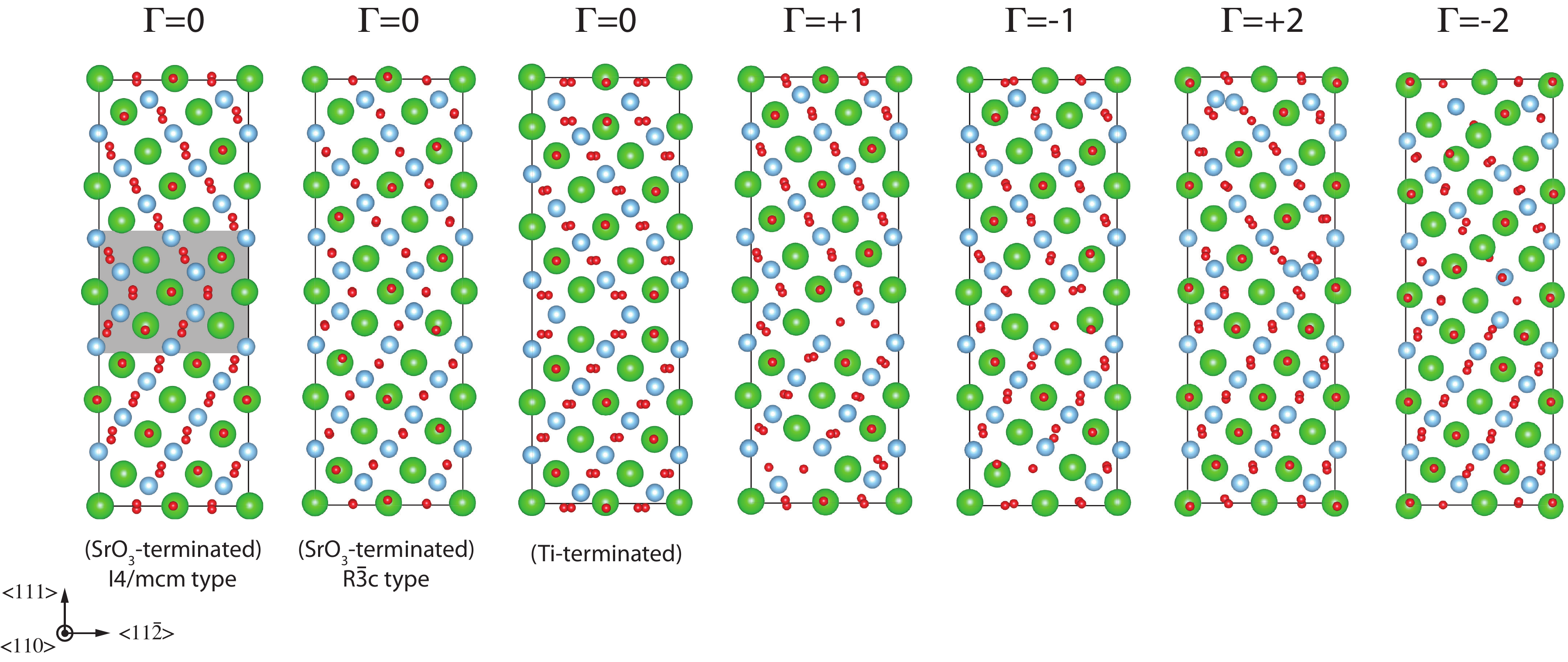}
\caption{
Predicted atomic structure of the  SrTiO$_3$ grain boundaries for stoichiometries $\Gamma=0, \pm 1, \pm 2$. Three structures for the stoichiometric condition ($\Gamma=0$) are shown, the two degenerate low energy SrO$_3$-terminated structures and the higher energy Ti-terminated structure. For the lowest energy structure for non-stoichiometric, SrO-rich conditions ($\Gamma=+1$), the two adjacent grains are sheared by $0.5\text{\AA}$ along the $\langle110\rangle$ direction. All structures show significant oxygen displacements from the bulk low temperature $I4/mcm$ phase. The randomization region used for interface prediction is indicated for the SrO$_3$-terminated $\Gamma=0$ structure in grey; all atoms in this region are randomized. The extent of the randomization region is $9.5\text{\AA} \times 9.2\text{\AA} \times 5.5\text{\AA}$ along $\langle 11\bar{2} \rangle$, $\langle 111 \rangle$ and $\langle 110 \rangle$  directions. All structures have two symmetric grain boundaries in the supercell due to periodicity. Red, green and blue circles represent O, Sr and Ti atoms, respectively.
}
\label{Fig:STO:GB:structures}
}
\end{figure*}

We next consider the SrTiO$_3$ system with a grain boundary. The setup for searching is shown in Fig.~\ref{Fig:STO:GB:structures}. The two crystals surrounding the randomization region are terminated each by $\left(111\right)$ planes, thereby biasing the system towards a $\Sigma 3 \left(111\right)$ type grain boundary. During searching a total number $N=29$ to $33$ of Sr, Ti and O atoms with different stoichiometry are added to the randomization region, surrounded by 88 atoms in the $\left(111\right)$-terminated grains. Structure prediction applied to this interface is significantly more challenging than for the graphene system. The number of atoms in the randomization region approximately doubles and three different atomic species have to be considered, increasing the search space significantly. We approximately enforce species distance constraints taken from the bulk compound. Many high energy structures are thereby eliminated, allowing us to sample the physically sensible search space more efficiently. Further complexity is added to the problem of performing structure prediction for SrTiO$_3$, since this system exhibits several different crystal phases that are very close in energy. For the bulk system we find at least three low energy bulk phases, $I4/mcm$, $R\bar{3}c$ and $Pm\bar{3}m$, in order of decreasing stability. The experimental tetragonal structure with $I4/mcm$ space group for temperatures $T<105\text{K}$ is reproduced as the ground state using the LDA for bulk SrTiO$_3$.

\begin{figure}
\center{
\includegraphics[width=8.6cm]{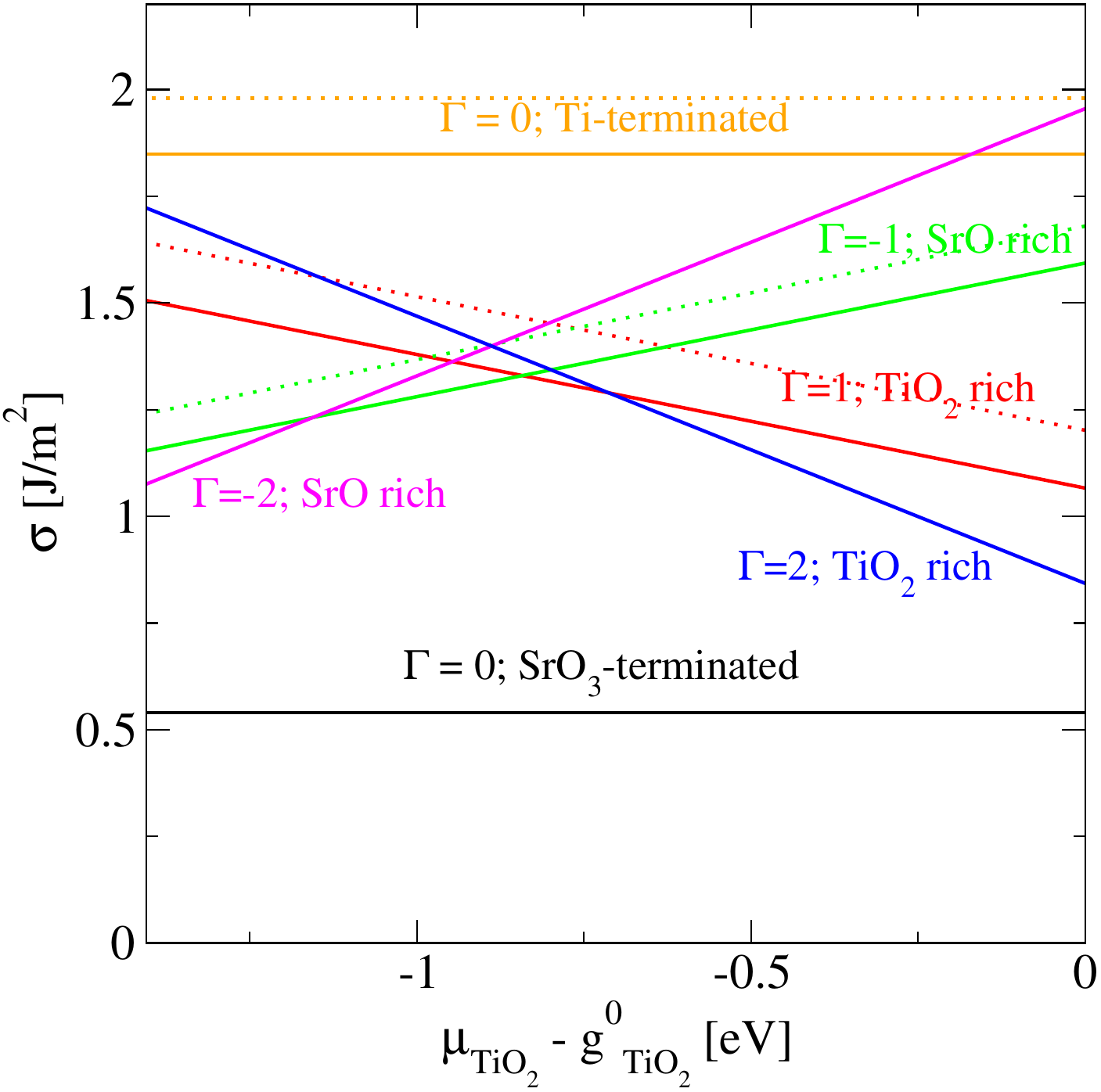}

\caption{
Grain boundary interface energy as a function of $\mu_{\text{TiO}_2}$, for the $\Sigma 3 \left(111\right)$ grain boundary in SrTiO$_3$. The interface energy for five different stoichiometries, $\Gamma=0; \pm1;\pm2$, are shown as solid lines for our results, with the overall lowest energy structure being the SrO$_3$-terminated stoichiometric structure. Also shown in dotted lines are the previously known lowest energy structures for $\Gamma=\pm1$ from Ref.~\onlinecite{GA:STO:GB:Finnis:Sutton:2010} and the ideal Ti-terminated structure in the Pm-3m phase.
}
\label{Fig:STO:GB:energy}
}
\end{figure}

We consider various stoichiometries, however we limit ourselves to adding or removing units of the binary compounds SrO and TiO$_2$. This simplifies the problem as it limits the search space to charge neutral configurations. We define the interface energy, $\sigma$, for the SrTiO$_3$ grain boundary with respect to the chemical potentials of its binary compounds SrO and TiO$_2$,
 \begin{equation}\label{Eq:STO:interface}
\sigma=\frac{1}{2A}\left(G_{tot}-n_{\text{SrO}}\mu_{\text{SrO}}-n_{\text{TiO}_2}\mu_{\text{TiO}_2}\right), 
\end{equation}
where G$_{tot}$ is the Gibbs free energy of the cell containing the grain boundary, $n_{x}$ is the total number of units of each binary compound, $x=\text{SrO, TiO}_2$ and $\mu_{x}$ is the chemical potential of each binary compound. The chemical potential for TiO$_2$ and SrO can only be determined to be within a range of $g_{\text{SrO}}^0 + \Delta G \leq \mu_{\text{SrO}} \leq  g_{SrO}^0 $ and $g_{\text{TiO}_2}^0 + \Delta G \leq \mu_{\text{TiO}_2} \leq  g_{\text{TiO}_2}^0 $, where $\Delta G$ is the formation energy of SrTiO$_3$ with respect to the binary compounds and $g_x^0$ is the free energy of the binary compounds in their ground state per formula unit.~\cite{Kaxiras:PRB:GaAs:1987, Qian:Martin:Chadi:PRB:1988, GA:STO:GB:Finnis:Sutton:2010} We treat SrTiO$_3$ in its low temperature $I4/mcm$ phase, SrO in its rocksalt and TiO$_2$ in its rutile structure. To consider either SrO or TiO$_2$ rich conditions it is convenient to write the above inequalities as   $\mu_{SrO}=g_{SrO}+\left(1-\lambda\right)\Delta G$ and $\mu_{TiO_2}=g_{TiO_2}+\lambda\Delta G$ with $0\leq\lambda\leq1$. This then allows one to express the interface energy by considering different stoichiometries  $\Gamma=n_{TiO_2}-n_{SrO}$ as,
 \begin{equation}\label{Eq:STO:interface:rearranged}
\sigma=\frac{1}{2A}\left[G_{tot}-n_{SrO}g_{SrTiO_3} - \Gamma \left(g_{TiO_2} +  \lambda \Delta G \right)\right]. 
\end{equation}
We approximate the Gibbs free energy by the respective total energies from DFT calculations. 

The stoichiometries we consider are for $\Gamma=0;\pm1;\pm2$. The interface formation energy, $\sigma$, for each stoichiometry is shown in Fig.~\ref{Fig:STO:GB:energy}. The atomic coordinates of the crystal structures for the lowest energy configurations of each stoichiometry are given as {\it cif} files in the Supplementary Material for completeness. We consider first the stoichiometric structures, where $\Gamma=0$. We find two primary structures, a SrO$_3$-terminated and a Ti-terminated grain boundary, with the lowest energy structure being the SrO$_3$ structure. Our Ti-terminated structure is significantly lower in energy than previous results for a Ti-terminated $\Sigma 3 \left(111\right)$ grain boundary in SrTiO$_3$.~\cite{GA:STO:GB:Finnis:Sutton:2010} The previous work found the structure to be in the $Pm\bar{3}m$ phase in the bulk, whereas our DFT-based search results show that the Ti-terminated structure assumes a lower energy distorted $I4/mcm$-type structure in the bulk part. We find the interface energy with the ideal $Pm\bar{3}m$ structure to be $\sigma^{\text{Pm}\bar{3}\text{m}}_{Ti}=1.98\text{~J/m}^2$ in close agreement with Ref.~\onlinecite{GA:STO:GB:Finnis:Sutton:2010}. In comparison, the oxygen distortions seen in Fig.~\ref{Fig:STO:GB:structures} lower the formation energy to $\sigma_{Ti}=1.85\text{~J/m}^2$.

The interface energy we find for the SrO$_3$-terminated structure is  $\sigma_{\text{SrO}_3}=0.54\text{~J/m}^2$, in agreement with previous results of $0.57\text{~J/m}^2$.~\cite{Benedek:Chua:Finnis:2008} We however find two degenerate structures for the SrO$_3$ interface, one distorted $I4/mcm$ phase, the other a distorted $R\bar{3}c$ phase. These distortions away from the bulk crystal phases extend far from the grain boundary. The strong distortions may significantly affect the material properties in the region of the grain boundary and warrants further investigation.

\begin{figure*}
\center{
\includegraphics[width=17.2cm, clip]{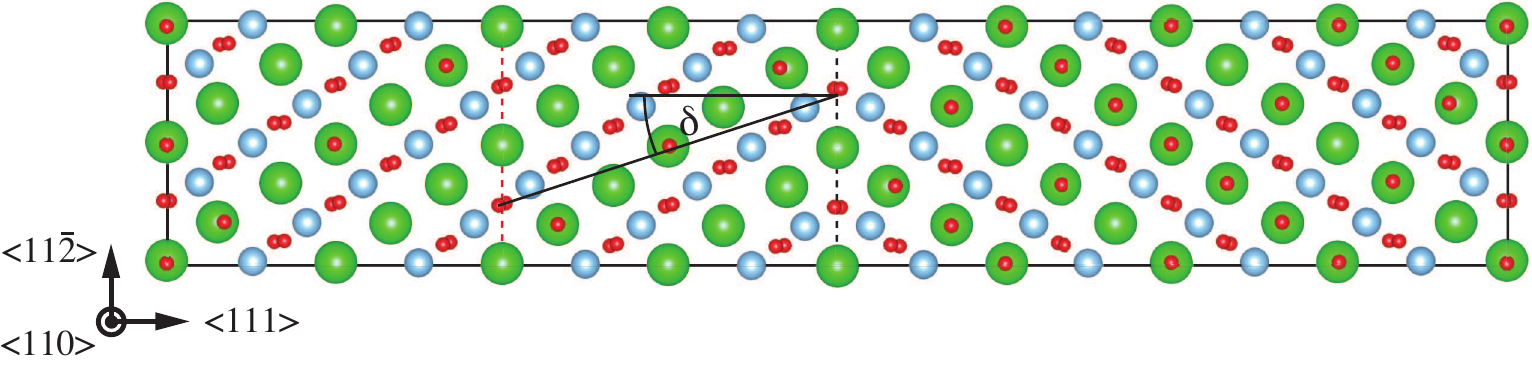}

\caption{
Stoichiometric ($\Gamma=0$) SrO$_3$-terminated SrTiO$_3$ grain boundary with 240 atoms in the unit cell. The crystal structure of the bulk part is of distorted $I4/mcm$ type. The distortions reach far into the bulk material. The angle $\delta$ that the vector between two O atoms makes with the normal of the grain boundary plane is shown for one set of O atoms at the mid-point between the two periodic grain boundaries. The mid-point between the grain boundaries is indicated by a dashed red line, whilst the center grain boundary plane is indicated by a dashed black line and its periodic image by solid vertical lines at the edges of the cell.  Red, green and blue circles represent O, Sr and Ti atoms, respectively.
}
\label{Fig:STO:GB:240:I4mcm}
}
\end{figure*}

\begin{figure}
\center{
\includegraphics[width=8.6cm, clip]{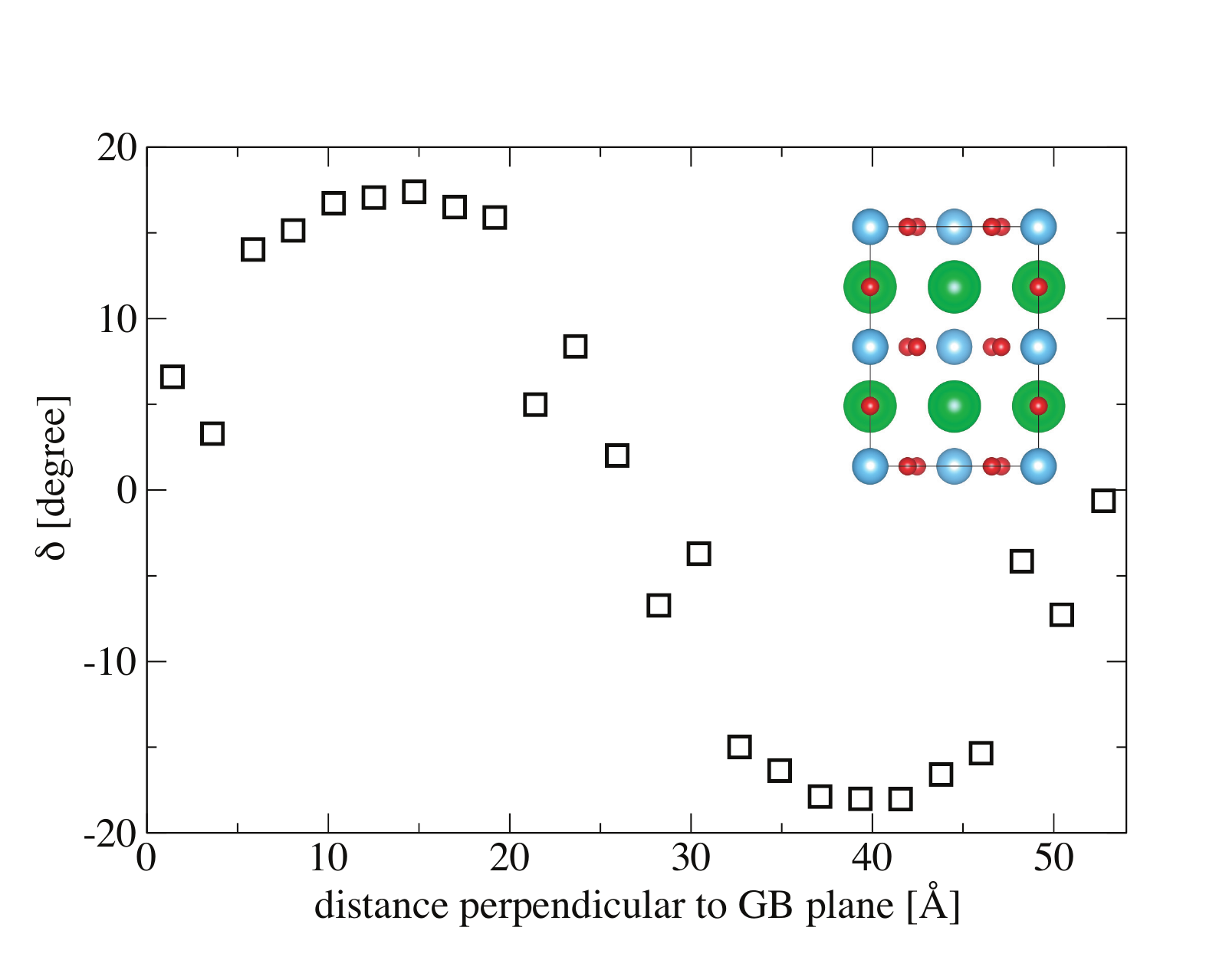}
\caption{
Distortion angle, $\delta$, as a function of distance perpendicular to the grain boundary of Fig.~\ref{Fig:STO:GB:240:I4mcm}. The distortions reach far into the bulk material, where at $14~\text{\AA}$ from the grain boundary a significant distortion of $\delta\sim20^{\circ}$ can still be observed, as opposed to $\delta^{0}=35.2^{\circ}$ for the bulk $I4/mcm$ bulk without a grain boundary defect. The ideal $I4/mcm$ structure is shown in the inset. 
}
\label{Fig:STO:GB:240:I4mcm:phase:angle}
}
\end{figure}

We have performed calculations with twice the unit cell perpendicular to the grain boundary plane (a total of 240 atoms), in order to investigate how far the distortions from the usual $I4/mcm$ ground state extend into the bulk. The fully relaxed structure is shown for the $I4/mcm$ type interface in Fig.~\ref{Fig:STO:GB:240:I4mcm}. In order to investigate the distortions from the ideal $I4/mcm$ structure we consider the angle, $\delta$, that the vector between pairs of two O atoms ({\it i.e.} pairs in the row of Ti and O atoms) projected onto the $\left(110\right)$ plane make with the normal of the GB plane. For the ideal $I4/mcm$ structure  this angle would be $\delta^{0}=35.2^{\circ}$. We see that in order to minimize distortions at the center plane of the grain boundary and to ensure matching of the two grains, the pairs of O atoms there align approximately parallel to the normal of the grain boundary. This angle does not fully recover to $35.2^{\circ}$, even with a supercell size of $54\text{\AA}$ (see Fig.~\ref{Fig:STO:GB:240:I4mcm} and \ref{Fig:STO:GB:240:I4mcm:phase:angle}). Since this structure has two periodic grain boundaries, this means that each grain boundary distorts the lattice over more than $14{~\text\AA}$, suggesting that the distortions are a very long-range effect. Care was taken to ensure that the structures were appropriately relaxed. The lattice constant perpendicular to the grain boundary is carefully relaxed. The lattice constants parallel to the interface were initially chosen to correspond to either the bulk $I4/mcm$, $R\bar{3}c$ or $Pm\bar{3}m$ phase lattice constants to simulate the bulk crystal structure far away from the grain boundary. In separate calculations we also fully relaxed the lattice constants parallel to the grain boundary to ensure no accidental bias towards one crystal phase. We further perform calculations where we double the cell size along $\langle11\bar{2}\rangle$ and $\langle110\rangle$, respectively, to ensure the periodicity does not constrain the system. The distortions and the energetic ordering of the structures remain the same for all cases.

We find that our structure prediction method is unbiased with respect to the crystal phase we initiate the system in. We have performed searches for which the two crystals surrounding the randomization region were initially in either the $I4/mcm$, $R\bar{3}c$ or $Pm\bar{3}m$ phase and consistently found SrO$_3$-terminated interfaces with the same distorted $I4/mcm$ or $R\bar{3}c$ structure.

We further consider 4 different non-stoichiometric conditions, $\Gamma=\pm 1$ and  $\Gamma=\pm 2$, where $\Gamma > 0$ is TiO$_2$ rich and $\Gamma < 0$ is SrO rich. The structures for $\Gamma=\pm2$ are found to be lower in energy than those for $\Gamma=\pm1$ for most values of the chemical potential of TiO$_2$, $\mu_{\text{TiO}_2}$. Our results for  $\Gamma=+1$ and  $\Gamma=-1$ are shown as solid red and green lines in Fig.~\ref{Fig:STO:GB:energy}, respectively. This is compared to previous results for the same stoichiometry shown as dotted red and green lines. For $\Gamma=-1$ we find a similar structure as in Ref.~\onlinecite{GA:STO:GB:Finnis:Sutton:2010}, however as for $\Gamma=0$ we find a structure with oxygen displacements that lower the energy in comparison to their $Pm\bar{3}m$ structures. The structure we find for $\Gamma=+1$ (shown in Fig.~\ref{Fig:STO:GB:structures}) is altogether different. In contrast to the structure from Chua {\it et al.},~\cite{GA:STO:GB:Finnis:Sutton:2010} we find that the two grains are sheared with respect to another by approximately $0.5\text{~\AA}$ along the $\langle110\rangle$ direction. The grain boundary structure at the interface is also significantly different, overall resulting in a lower energy. 

Although most of our results give symmetric structures with no significant shear of the two grains, the constraints we impose do not prohibit the system from reaching such structures. Many high energy structures were in fact found that had significant shear; instead we conclude that the $\Sigma 3 \left(111\right)$ GB merely energetically prefers configurations with little or no shear, and only find a small shear displacement for $\Gamma=+1$.

\section{Conclusion}\label{Sec:Conclusion}

We have shown how {\it ab initio} random structure searching can be used to study interfaces with variable stoichiometry and have found new low energy structures for both the graphene and SrTiO$_3$ grain boundary systems. Previous work on structure prediction of the graphene and SrTiO$_3$ grain boundaries have missed several important structural details for the low energy configurations. It is not clear if this is due to the different searching algorithms employed, {\it i.e.} random structure searching as opposed to evolutionary algorithms, or due to searching with classical interatomic potentials instead of DFT. It is important to note however that the ground state found by searching with a classical potential followed by evaluation of the resulting structures with DFT is inherently not the ground state structure of DFT but instead such a procedure only gives a more accurate value for the energy of the ground state of the classical potential. Moreover, we show in this work that treating system sizes previously only studied with structure prediction methods based on classical potentials, are now well within the reach of treatment with DFT in combination with an efficient searching algorithm and appropriate constraints. Our method is unbiased with respect to the initial crystal phase of the grains surrounding the randomization region and able to find subtle structural details in the bulk caused by the presence of the grain boundary: We find for the SrTiO$_3$ grain boundary that structures with long-range distortions due to the grain boundary lower the interface energy even for stoichiometric conditions, whilst the genetic algorithm using a classical interatomic potential used in Ref.~\onlinecite{GA:STO:GB:Finnis:Sutton:2010} predicted all structures to be in the $Pm\bar{3}m$ phase. At the same time we are able to treat variable stoichiometry and by virtue of being first-principles and not requiring any parametrization or system-specific interatomic potentials, our method can be easily applied to other materials systems without the need to alter our approach. These are all crucial aspects of any method attempting to address the emerging field of interface discovery. Advances in the structure prediction of interfaces will increase our understanding of the interface structure/property relation of polycrystalline and heterostructure materials, which in turn will open the possibility to develop materials with specific interfaces that give them desired properties.

This work was supported in part by the EPSRC Grant EP/G007489/2. We thank Nicole Benedek for useful discussions and for providing their structure files of the SrTiO$_3$ grain boundary from Ref.~\onlinecite{GA:STO:GB:Finnis:Sutton:2010}.  Computational resources from the University College London and London Centre for Nanotechnology Computing Services as well as HECToR and Archer as part of the UKCP consortium are gratefully acknowledged.

\end{document}